\documentclass[twoside,twocolumn,9pt]{article}
\usepackage{extsizes}
\usepackage[super,sort&compress,comma]{natbib} 
\usepackage[version=3]{mhchem}
\usepackage[left=1.5cm, right=1.5cm, top=1.785cm, bottom=2.0cm]{geometry}
\usepackage{balance}
\usepackage{mathptmx}
\usepackage{sectsty}
\usepackage{graphicx} 
\usepackage{lastpage}
\usepackage[format=plain,justification=justified,singlelinecheck=false,font={stretch=1.125,small,sf},labelfont=bf,labelsep=space]{caption}
\usepackage{fancyhdr}
\usepackage{fnpos}
\usepackage[english]{babel}
\addto{\captionsenglish}{%
  
}
\usepackage{array}
\usepackage{droidsans}
\usepackage{charter}
\usepackage[T1]{fontenc}
\usepackage[usenames,dvipsnames]{xcolor}
\usepackage{setspace}
\usepackage[compact]{titlesec}
\usepackage{hyperref}

\usepackage{epstopdf}

\usepackage[mathscr]{euscript}
\usepackage{xcolor}
\usepackage{xspace}
\usepackage{amsmath}
\usepackage{dsfont}
\usepackage{placeins}
\usepackage{soul}
\usepackage{cuted}
\usepackage{MnSymbol}%
\usepackage{wasysym}
\DeclareMathOperator{\arctantwo}{arctan2}
\newcommand{\blue}[1]{\textcolor{black}{#1}}

\newcommand{\degC}{\ensuremath{{}^{\circ}\mathrm{C}}\xspace}
\newcommand{\uT}{\ensuremath{\,\mu\mathrm{T}}\xspace}
\newcommand{\uL}{\ensuremath{\,\mu\mathrm{L}}\xspace}

\DeclareMathAlphabet{\mymathbb}{U}{BOONDOX-ds}{m}{n}

\def\Cth{\ensuremath{{}^{13}\mathrm{C}}\xspace}

\def\Hone{\ensuremath{{}^{1}\mathrm{H}}\xspace}

\newcommand{\RuCat}{$\mathrm{[RuCp^\ast(MeCN)_3]PF_6}$\xspace}
\newcommand{\iu}{{i\mkern1mu}}

\newcommand\ket[1]{|#1\rangle}
\newcommand\bra[1]{\langle#1|}

\newcommand{\hf}{\tfrac{1}{2}}
\newcommand{\para}{\emph{para}\xspace}
\newcommand{\Para}{\emph{Para}\xspace}
\newcommand{\Htwo}{\ensuremath{\mathrm{H_2}}\xspace}

\newcommand{\Bbias}{B_\mathrm{bias}}

\newcommand{\BSTORM}{B_{\rm STORM}}
\newcommand{\gI}{\gamma_I}
\newcommand{\gS}{\gamma_S}

\newcommand{\HJ}{H_J}

\newcommand{\boldI}{{\bf I}}
\newcommand{\boldS}{{\bf S}}

\newcommand{\wSTORM}{\omega_\mathrm{STORM}}
\newcommand{\wST}{\omega_{ST}}

\newcommand{\wSTnut}{\omega_{ST}^\mathrm{nut}}
\newcommand{\wrot}{\omega_{\rm rot}}
\newcommand{\wLI}{\omega^{I}_{0}}
\newcommand{\wLS}{\omega^{S}_{0}}
\newcommand{\wRI}{\omega^{I}_{1}}
\newcommand{\wRS}{\omega^{S}_{1}}
\newcommand{\weff}{\omega_{\rm eff}}
\newcommand{\theff}{\theta_{\rm eff}}

\definecolor{cream}{RGB}{222,217,201}

\begin{document}

\pagestyle{fancy}
\thispagestyle{plain}
\fancypagestyle{plain}{
\renewcommand{\headrulewidth}{0pt}
}

\makeFNbottom
\makeatletter
\renewcommand\LARGE{\@setfontsize\LARGE{15pt}{17}}
\renewcommand\Large{\@setfontsize\Large{12pt}{14}}
\renewcommand\large{\@setfontsize\large{10pt}{12}}
\renewcommand\footnotesize{\@setfontsize\footnotesize{7pt}{10}}
\makeatother

\renewcommand{\thefootnote}{\fnsymbol{footnote}}
\renewcommand\footnoterule{\vspace*{1pt}%
\color{cream}\hrule width 3.5in height 0.4pt \color{black}\vspace*{5pt}} 
\setcounter{secnumdepth}{5}

\makeatletter 
\renewcommand\@biblabel[1]{#1}            
\renewcommand\@makefntext[1]%
{\noindent\makebox[0pt][r]{\@thefnmark\,}#1}
\makeatother 
\renewcommand{\figurename}{\small{Fig.}~}
\sectionfont{\sffamily\Large}
\subsectionfont{\normalsize}
\subsubsectionfont{\bf}
\setstretch{1.125} 
\setlength{\skip\footins}{0.8cm}
\setlength{\footnotesep}{0.25cm}
\setlength{\jot}{10pt}
\titlespacing*{\section}{0pt}{4pt}{4pt}
\titlespacing*{\subsection}{0pt}{15pt}{1pt}

\fancyfoot{}
\fancyfoot[LO,RE]{\vspace{-7.1pt}\includegraphics[height=9pt]{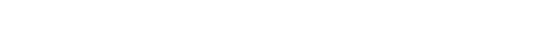}}
\fancyfoot[CO]{\vspace{-7.1pt}\hspace{11.9cm}\includegraphics{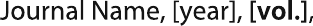}}
\fancyfoot[CE]{\vspace{-7.2pt}\hspace{-13.2cm}\includegraphics{head_foot/RF}}
\fancyfoot[RO]{\footnotesize{\sffamily{1--\pageref{LastPage} ~\textbar  \hspace{2pt}\thepage}}}
\fancyfoot[LE]{\footnotesize{\sffamily{\thepage~\textbar\hspace{4.65cm} 1--\pageref{LastPage}}}}
\fancyhead{}
\renewcommand{\headrulewidth}{0pt} 
\renewcommand{\footrulewidth}{0pt}
\setlength{\arrayrulewidth}{1pt}
\setlength{\columnsep}{6.5mm}
\setlength\bibsep{1pt}

\makeatletter 
\newlength{\figrulesep} 
\setlength{\figrulesep}{0.5\textfloatsep} 

\newcommand{\topfigrule}{\vspace*{-1pt}%
\noindent{\color{cream}\rule[-\figrulesep]{\columnwidth}{1.5pt}} }

\newcommand{\botfigrule}{\vspace*{-2pt}%
\noindent{\color{cream}\rule[\figrulesep]{\columnwidth}{1.5pt}} }

\newcommand{\dblfigrule}{\vspace*{-1pt}%
\noindent{\color{cream}\rule[-\figrulesep]{\textwidth}{1.5pt}} }

\makeatother

\twocolumn[
  \begin{@twocolumnfalse}
{\hfill\raisebox{0pt}[0pt][0pt]{\includegraphics[height=55pt]{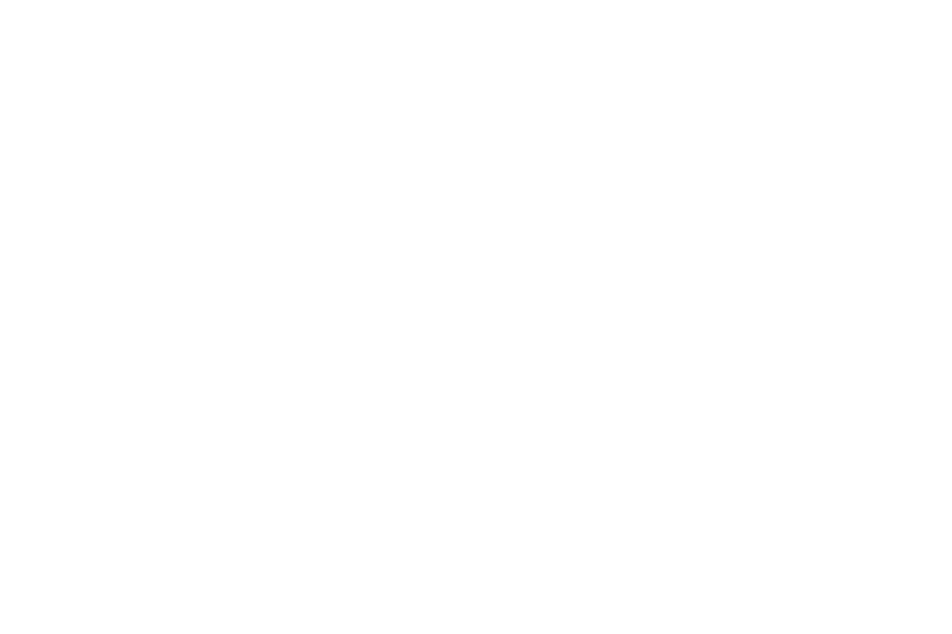}}\\[1ex]}\par
\vspace{1em}
\sffamily
\begin{tabular}{m{4.5cm} p{13.25cm} }

\includegraphics{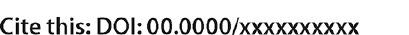} & \noindent\LARGE{\textbf{Hyperpolarization read-out through rapidly rotating fields in the zero- and low-field regime}} \\
\vspace{0.3cm} & \vspace{0.3cm} \\

 & \noindent\large{Laurynas Dagys,$^{\ast}$\textit{$^{a}$} and Christian Bengs\textit{$^{a}$}} \\

\includegraphics{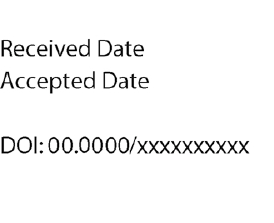} & \noindent\normalsize{An integral part of \para-hydrogen induced polarization (PHIP) methods is the conversion of nuclear singlet order into observable magnetization.  In this study polarisation transfer to a heteronucleus is achieved through a selective rotation of the proton singlet-triplet states driven by a combination of a rotating magnetic field and a weak bias field.  
Surprisingly we find that efficient polarisation transfer driven by a STORM (Singlet-Triplet Oscillations through Rotating Magnetic fields) pulse in the presence of sub-$\mu$T bias fields requires rotation frequencies on the order of several kHz. The rotation frequencies therefore greatly exceed any of the internal frequencies of typical zero- to ultralow field experiments. We further show that the rotational direction of the rotating field is not arbitrary and greatly influences the final transfer efficiency.
Some of these aspects are demonstrated experimentally by considering hyperpolarised  \mbox{(1-\Cth)fumarate}. In addition, we provide numerical simulations highlighting the resilience of the STORM pulse against disruptive quadrupolar coupling partners. In contrast to most of the existing methods, the STORM procedure therefore represents a promising candidate for quadrupolar decoupled polarisation transfer in PHIP experiments.} \\

\end{tabular}

 \end{@twocolumnfalse} \vspace{0.6cm}

  ]

\renewcommand*\rmdefault{bch}\normalfont\upshape
\rmfamily
\section*{}
\vspace{-1cm}


\footnotetext{\textit{School of Chemistry, Highfield Campus, Southampton, United Kingdom, SO171BJ. E-mail: l.dagys@soton.ac.uk}}


\section{Introduction}
\label{sec:introduction}

The inherently low sensitivity of Nuclear Magnetic Resonance (NMR) may be greatly overcome through the use of hyperpolarization methods~\cite{natterer_parahydrogen_1997, bowers_parahydrogen_1987,ardenkjaer-larsen_increase_2003,maly_dynamic_2008,walker_spin-exchange_1997,kovtunov_hyperpolarized_2018, emondts_polarization_2017,stephan_13c_2002,adams_reversible_2009,johannesson_transfer_2004,theis_light-sabre_2014,theis_microtesla_2015,cavallari_effects_2015,eills_polarization_2019,bengs_robust_2020,devience_homonuclear_nodate,rodin_constant-adiabaticity_2021-1,rodin_constant-adiabaticity_2021,knecht_rapid_2021,dagys_low-frequency_2021,dagys_nuclear_2021}. At the core of these methods is the production of nuclear spin order far from thermal equilibrium that can lead to signal enhancements of many orders of magnitude.

Particular promising techniques are \para-hydrogen induced polarisation (PHIP) methods~\cite{natterer_parahydrogen_1997,bowers_parahydrogen_1987,emondts_polarization_2017,stephan_13c_2002,adams_reversible_2009,johannesson_transfer_2004,theis_light-sabre_2014,theis_microtesla_2015,cavallari_effects_2015,eills_polarization_2019,bengs_robust_2020,devience_homonuclear_nodate,rodin_constant-adiabaticity_2021-1,rodin_constant-adiabaticity_2021,knecht_rapid_2021,dagys_low-frequency_2021,dagys_nuclear_2021}. Methods of this type utilise molecular hydrogen gas enriched in its \para-spin isomer, typically achieved by passing the cooled gas over an iron oxide catalyst~\cite{bowers_parahydrogen_1987,natterer_parahydrogen_1997}. 

For the case of hydrogenative-PHIP (considered here) the \para-enriched hydrogen gas (\para-hydrogen) is allowed to react with a suitable precursor molecule. Upon hydrogenation the nuclear singlet order of the hydrogen gas is carried over to the product molecule. However, the resulting nuclear singlet order located on the product molecule is NMR silent. Efficient conversion of nuclear singlet order into observable magnetization is thus an integral part of the method.

A number of techniques already exist for this purpose, both at high and low magnetic fields~\cite{johannesson_transfer_2004,theis_light-sabre_2014,theis_microtesla_2015,cavallari_effects_2015,eills_polarization_2019,bengs_robust_2020,devience_homonuclear_nodate,rodin_constant-adiabaticity_2021-1,rodin_constant-adiabaticity_2021,knecht_rapid_2021,dagys_low-frequency_2021,dagys_nuclear_2021}. High field methods benefit from the usual advantages, spectral separation between hetero- and homonuclei, strong pulse schemes with error compensation and applicability to a broad class of molecular systems~\cite{devience_preparation_2013,pravdivtsev_highly_2014,bengs_robust_2020,dagys_nuclear_2021,theis_light-sabre_2014,haake_efficient_1996}. However, these benefits often come at a price. Some technical challenges may arise due to additional relaxation phenomena and coherent leakage, which may lead to significant polarization losses~\cite{rodin_constant-adiabaticity_2021-1,knecht_mechanism_2018, knecht_indirect_2019}.

Some of these issues may be circumvented at low magnetic fields~\cite{johannesson_transfer_2004,cavallari_effects_2015,eills_polarization_2019,rodin_constant-adiabaticity_2021,rodin_constant-adiabaticity_2021-1,knecht_rapid_2021}. This has been utilised to produce large quantities of chemically pure and hyperpolarized (1-\Cth)fumarate, for example~\cite{knecht_rapid_2021}. The reaction  was carried out inside a pressurised metal reactor, which itself was placed inside a magnetic shield. The polarization transfer was performed by sweeping the magnetic field in the sub-microtesla regime. Such a setup would be impossible at high magnetic fields as the reaction vessel is incompatible with pulsed radio-frequency methods. 

We have recently demonstrated that efficient polarisation transfer may also be performed in the presence of a weak static magnetic field superimposed with a weak oscillating low field (WOLF) along the same direction~\cite{dagys_low-frequency_2021}. A magnetic field geometry with the oscillating field applied along the same direction as the main magnetic field is unusual for NMR, indeed if the oscillating field is applied in the conventional transverse plane the WOLF pulse becomes ineffective.

Generally speaking, the description of oscillating fields in the low magnetic field regime is complicated. In contrast to high field experiments both the resonant part and the counter-rotating part of the linearly polarised field have to be considered~\cite{meriles_high-resolution_2004, sakellariou_nmr_2005}. However, \blue{at low magnetic fields it is technically trivial} to generate rotating magnetic fields that are resonant with the nuclear spin transition frequencies. This way perturbations due to the counter-rotating components are simply avoided.

In this work we demonstrate that the application of transverse rotating fields may be exploited for the polarisation transfer step in PHIP experiments. The application of a suitable STORM (Singlet-Triplet Oscillations through Rotating Magnetic fields) pulse enables polarisation transfer from the singlet pair to a heteronucleus leading to substantially enhanced NMR signals. We validate some of these concepts experimentally by generating hyperpolarized (1-\Cth)fumarate, and explore the STORM condition as a function of the bias field, rotation frequency and sense of rotation.

We find that driving spin transitions with a rotating magnetic field in low magnetic fields requires unusually high rotation frequencies, sometimes several kHz. This is in contrast to typical zero-to-ultralow field experiments which involve frequencies on the order of several Hz at most~\cite{pileio_extremely_2009,sjolander_13c-decoupled_2017,sjolander_transition-selective_2016}. The observed polarization levels are comparable to other low-field techniques~\cite{rodin_constant-adiabaticity_2021,rodin_constant-adiabaticity_2021-1,knecht_rapid_2021,dagys_low-frequency_2021}. However, we provide numerical evidence that the STORM method greatly outperforms several existing techniques in the presence of fast relaxing quadrupolar nuclei, such as deuterium for example. The STORM method might therefore represent a simple solution to quadrupolar decoupled polarisation transfer at low magnetic fields~\cite{birchall_quantifying_2020,tayler_scalar_2019}.

\section{Theory}
\label{sec:theory}

Consider an ensemble of nuclear three-spin-1/2 systems consisting of two nuclei of isotopic type $I$ and a third nucleus of isotopic type $S$. The nuclei are characterised by the magnetogyric ratio's $\gI$ and $\gS$, respectively. 
For an isotropic solution, the nuclei mutually interact by scalar spin-spin coupling terms
\begin{equation}
\begin{aligned}
    H_{J} = H_{II}+H_{IS},
    \end{aligned}
\end{equation}
where the Hamiltonian $H_{II}$ describes the homonuclear couplings
\begin{equation}
\begin{aligned}
    H_{II}=2\pi J_{12}\boldI_1\cdot\boldI_2
    \end{aligned}
\end{equation}
and the Hamiltonian $H_{IS}$ describes the heteronuclear couplings 
\begin{equation}
\begin{aligned}
    H_{IS}=
   2\pi J_{13}\boldI_1\cdot\boldS
   +2\pi J_{23}\boldI_2\cdot\boldS.
    \end{aligned}
\end{equation}
For the remainder of the discussion we assume the coupling constants $J_{13}$ and $J_{23}$ to be different ($J_{13}\neq J_{23}$) and \blue{a positive homonuclear coupling constant $J_{12}$. The heteronuclear $J$ coupling Hamiltonian may be split} in its symmetric and anti-symmetric part
\begin{equation}
\begin{aligned}
    H_{IS}=H^{\Sigma}_{IS}+H^{\Delta}_{IS},
    \end{aligned}
\end{equation}
with
\begin{equation}
\begin{aligned}
    &H^{\Sigma}_{IS}=\pi(J_{13}+J_{23})(\boldI_1+\boldI_2)\cdot \boldS,
    \\
    &H^{\Delta}_{IS}=\pi(J_{13}-J_{23})(\boldI_1-\boldI_2)\cdot \boldS.
    \end{aligned}
\end{equation}

\begin{figure}[h!]
\includegraphics[width=0.48\textwidth]{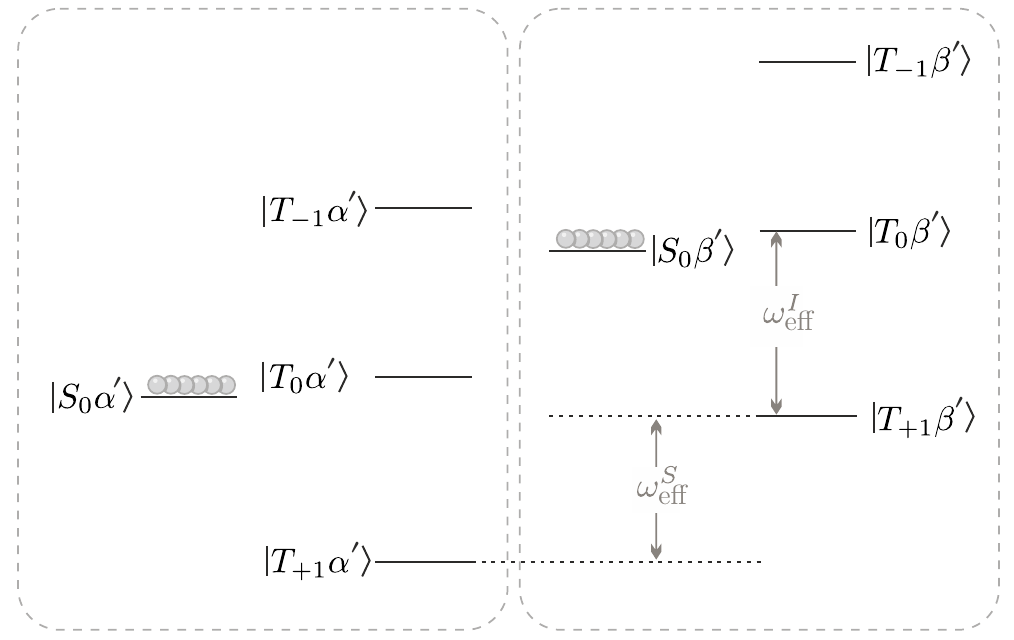}
\caption{\label{fig:energies} Eigenvalues and eigenstates of the effective field Hamiltonian (equation~\ref{eq:H_eff}) in the near-equivalence limit ($|J_{13}-J_{23}|\ll|J_{12}|$\blue{, with $J_{12}$ referring to the homonuclear $J$-coupling) and referring $J_{13}$, $J_{23}$ to the out-of-pair couplings.}  The circles represent the population distribution for a fully populated singlet state between the two $I$-spins. The effective nutation frequency $\omega_\mathrm{eff}^X$ is given by equation~\ref{eq:omega_eff} where $X$ refers to either spin \textit{I} or \textit{S}. \blue{Primes indicate alignment of the nuclear spins states along the effective field.}
}
\end{figure}

The nuclear spin ensemble may further be manipulated by the application of external magnetic fields. The magnetic field Hamiltonian is constructed by coupling the spin angular momenta to the external magnetic field taking their respective magnetogyric ratio's into account
\begin{equation}
\begin{aligned}
 H_{M}(t)=-\gI {\bf B}(t)\cdot(\boldI_{1}+\boldI_{2})-\gS {\bf B}(t)\cdot\boldS.
    \end{aligned}
\end{equation}
The total spin Hamiltonian is then given by
\begin{equation}
\begin{aligned}
H(t)=H_{J}+H_{M}(t).
    \end{aligned}
\end{equation}

\subsection{Rotating field Hamiltonian}

\begin{figure}[h!]
\includegraphics[width=0.48\textwidth]{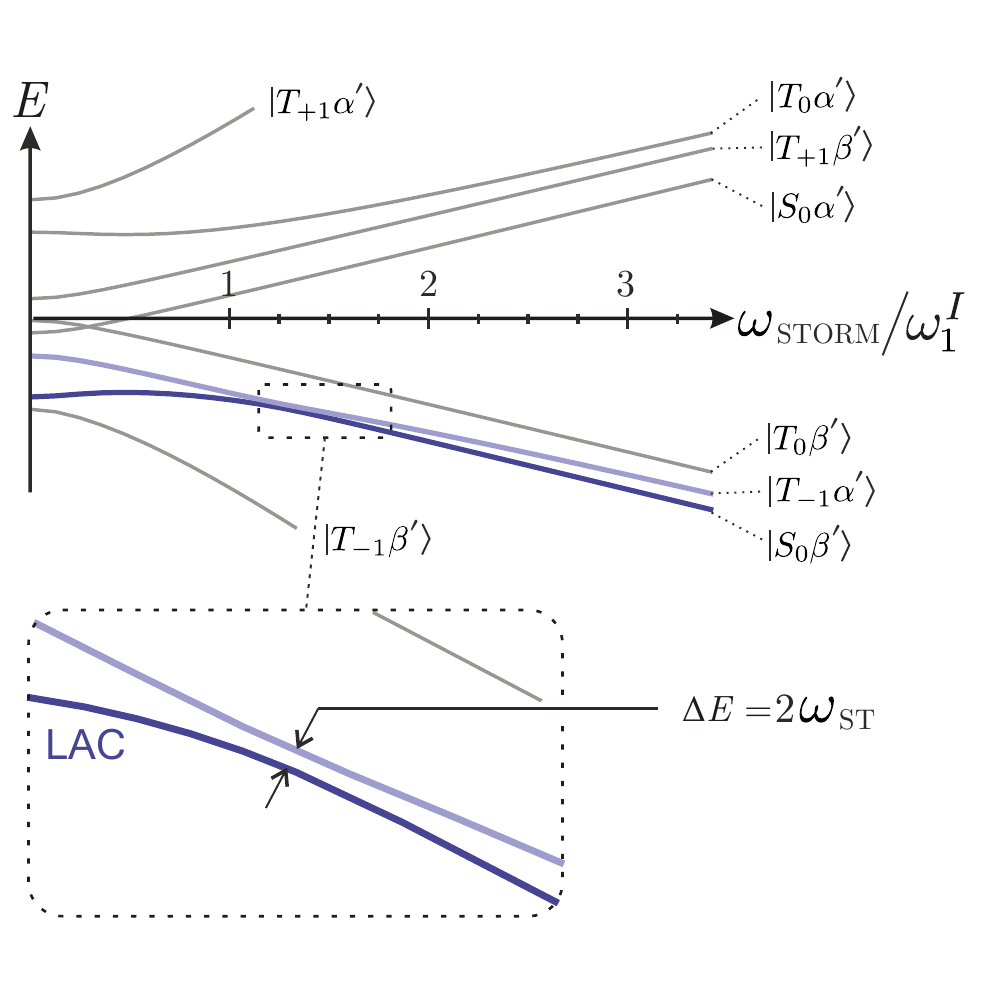}
\caption{\label{fig:lac} 
Eigenvalues and eigenstates of the interaction frame Hamiltonian (equation~\ref{eq:H_tilde}) as a function of the rotating magnetic field frequency $\wrot$ at zero-field. The states $\ket{S_{0}\beta^{'}}$ and $\ket{T_{-1}\alpha^{'}}$ undergo a Level-Anti Crossing (LAC) highlighted by the inset. \blue{Note that $\omega_1^I/\omega_1^S\approx4$ for $I=^{1}$H and $S=^{13}$C.} 
}
\end{figure}

Consider now the application of a time-dependent rotating magnetic field in the presence of a weak bias field along the laboratory frame $z$-axis. The $z$-bias Hamiltonian is given by
\begin{equation}
\begin{aligned}
H_{\rm bias}
&=-\gI B_{\rm bias} (I_{1z}+I_{2z})-\gS B_{\rm bias} S_{z}
\\
&=\wLI(I_{1z}+I_{2z})+\wLS S_{z},
    \end{aligned}
\end{equation}
whereas the rotating magnetic field Hamiltonian is given by
\begin{equation}
\begin{aligned}
H_{\rm rot}(t)
=&B_{\rm rot} \cos(\wrot t)(-\gI (I_{1x}+I_{2x})-\gS S_{x})
\\
-&B_{\rm rot} \sin(\wrot t)(-\gI (I_{1y}+I_{2y})-\gS S_{y}).
    \end{aligned}
\end{equation}
The total spin Hamiltonian may now be expressed as a combination of scalar-coupling terms, the bias term and the rotating field contribution
\begin{equation}
\begin{aligned}
H(t)=H_{J}+H_{M}(t)=H_{J}+H_{\rm bias}+H_{\rm rot}(t).
    \end{aligned}
\end{equation}

It turns out to be advantageous to isolate the rotating part of $H(t)$. To this end we consider an interaction frame transformation rotating all three spins equally around the laboratory frame $z$-axis. The angular frequency is chosen to coincide with $\wrot$
\begin{equation}
\begin{aligned}
K_{z}(t)=\exp\{-\iu (I_{1z}+I_{2z}+S_{z})\wrot t\}.
    \end{aligned}
\end{equation}
The corresponding interaction frame Hamiltonian $\tilde{H}(t)$ is given by
\begin{equation}
\begin{aligned}
\tilde{H}=
&K_{z}(t)H(t)K_{z}(-t)+\iu \dot{K}_{z}(t)K_{z}(-t)
\\
=&\HJ+\wRI(I_{1x}+I_{2x})+\wRS S_{x}
\\
+&(\wLI+\wrot)(I_{1z}+I_{2z})+(\wLS+\wrot)S_{z},
    \end{aligned}
\end{equation}
which has the advantage of being time-independent.

\subsection{Effective field Hamiltonian}

Within the interaction frame the spins evolve under a new effective magnetic field $B_{\rm eff}$. The coupling of the effective field to the $I$ and $S$ spins may be characterised by the effective nutation frequencies $\weff^{X}$
\begin{equation}
\label{eq:omega_eff}
\begin{aligned}
&\weff^{X}=\sqrt{(\omega^{X}_{0}+\wrot)^{2}+(\omega^{X}_{1})^{2}},
    \end{aligned}
\end{equation}
and the polar angles $\theff^{X}$
\begin{equation}
\begin{aligned}
&\theff^{X}=\arctantwo(\omega^{X}_{0}+\wrot,\omega^{X}_{1}).
    \end{aligned}
\end{equation}
The polar angles describe the field direction with respect to the laboratory frame $z$-axis. An alternative representation of $\tilde{H}$ is thus given by
\begin{equation}
\label{eq:H_tilde}
\begin{aligned}
\tilde{H}= XH_{\rm eff}X^{\dagger}+H^{\Delta}_{IS},
    \end{aligned}
\end{equation}
where $H_{\rm eff}$ represents the effective field Hamiltonian in the absence of the anti-symmetric heteronuclear $J$-couplings
\begin{equation}
\label{eq:H_eff}
\begin{aligned}
H_{\rm eff}= \weff^{I}(I_{1z}+I_{2z})+\weff^{S}S_{z}+H_{II}+H^{\Sigma}_{IS}.
    \end{aligned}
\end{equation}
The transformation $X$ is defined as a composite rotation of spins $I$ and $S$
\begin{equation}
\begin{aligned}
X=R^{12}_{y}(\theff^{I})R^{3}_{y}(\theff^{S}).
    \end{aligned}
\end{equation}

Consider now the set of STZ states aligned along the effective magnetic field direction
\begin{equation}
\begin{aligned}
&\ket{T_{m}\mu^{'}}=X\ket{T_{m}\mu},
\\
&\ket{S_{0}\mu^{'}}=X\ket{S_{0}\mu}.
    \end{aligned}
\end{equation}
These states are exact eigenstates of the effective field part of the interaction frame Hamiltonian
\begin{equation}
\begin{aligned}
XH_{\rm eff} X^{\dagger}\ket{T_{m}\mu^{'}}
=&XH_{\rm eff} X^{\dagger}X\ket{T_{m}\mu}
=\lambda X\ket{T_{m}\mu}
=\lambda\ket{T_{m}\mu^{'}},
    \end{aligned}
\end{equation}
and similarly for $\ket{S_{0}\mu^{'}}$. The rotated STZ states thus represent the approximate eigenstates of $\tilde{H}$ and the heteronuclear coupling term $H^{\Delta}_{IS}$ may be considered a perturbation. 

\subsection{Para-hydrogen Induced Polarisation}
\label{sec:PHIP}

For typical PHIP experiments involving $I_{2}S$ systems at sufficiently low magnetic fields we may approximate the initial state of the spin ensemble by pure singlet population
\begin{equation}
\begin{aligned}
\rho(0)=\ket{S_{0}}\bra{S_{0}}\otimes \frac{1}{2}\mathds{1}_{S},
    \end{aligned}
\end{equation}
where $\mathds{1}_{S}$ represents the unity operator for spin $S$. Because the initial state of the ensemble is rotationally invariant we may express $\rho(0)$ in a straightforward manner in the rotated STZ basis
\begin{equation}
\label{eq:rho0}
\begin{aligned}
\rho(0)=\frac{1}{2}\left(\ket{S_{0}\alpha^{'}}\bra{S_{0}\alpha^{'}}+\ket{S_{0}\beta^{'}}\bra{S_{0}\beta^{'}}\right).
    \end{aligned}
\end{equation}

In the absence of $H^{\Delta}_{IS}$ no heteronuclear magnetisation may be extracted out of the system, however the presence of $H^{\Delta}_{IS}$ causes coherent mixing within the manifolds $\{\ket{S_{0}\beta^{'}},\ket{T_{0}\alpha^{'}}, \ket{T_{-1}\alpha^{'}}\}$ and $\{\ket{S_{0}\alpha^{'}},\ket{T_{0}\beta^{'}}, \ket{T_{+1}\beta^{'}}\}$.

Strictly speaking these two manifolds are not completely isolated from all other states. But as illustrated in figure \ref{fig:lac} mixing of this type will be efficiently suppressed for our choice of the rotation frequency $\wrot$. Taking the first manifold for example, one may show to first order in perturbation theory that the following inequalities are well satisfied
\begin{equation}
\label{eq:TLS_cond}
\begin{aligned}
&\vert\bra{T_{-1}\alpha^{'}}\tilde{H}\ket{n}\vert\ll\vert\bra{T_{-1}\alpha^{'}}\tilde{H}\ket{T_{-1}\alpha^{'}}-\bra{n}\tilde{H}\ket{n}\vert,
\\
&\vert\bra{T_{0}\alpha^{'}}\tilde{H}\ket{n}\vert\;\;\ll\vert\bra{T_{0}\alpha^{'}}\tilde{H}\ket{T_{0}\alpha^{'}}-\bra{n}\tilde{H}\ket{n}\vert,
\\
&\vert\bra{S_{0}\beta^{'}}\tilde{H}\ket{n}\vert\;\;\ll\vert\bra{S_{0}\beta^{'}}\tilde{H}\ket{S_{0}\beta^{'}}-\bra{n}\tilde{H}\ket{n}\vert,
    \end{aligned}
\end{equation}
where $\ket{n}$ represents any state outside the manifold.

The energy separation between the $\ket{S_{0}\beta^{'}}$ and $\ket{T_{-1}\alpha^{'}}$ state is given by
\begin{equation}
\label{eq:delta_E}
\begin{aligned}
\Delta E
=\weff^{I}-\weff^{S}+2\pi\left(J_{12}-\frac{J_{13}+J_{23}}{4}\cos(\theff^{I}-\theff^{S})\right).
    \end{aligned}
\end{equation}
As a result, coherent state mixing is maximised by choosing an optimised rotation frequency $\wSTORM$ such that 
\begin{equation}
\begin{aligned}
\Delta E=0 \quad{\rm at}\quad\wrot=\wSTORM.
    \end{aligned}
\end{equation}
We refer to such a scenario as the application of a STORM pulse, which causes a degeneracy between the $\ket{S_{0}\beta^{'}}$ and $\ket{T_{-1}\alpha^{'}}$ state, and leads to a level anti-crossing (LAC) if the anti-symmetric heteronuclear $J$-couplings are included (see inset figure \ref{fig:lac})~\cite{rodin_representation_2020,messiah_albert_quantum_1962}. 

As shown in the appendix 
mixing between the $\ket{S_{0}\beta^{'}}$ and $\ket{T_{-1}\alpha^{'}}$ state occurs with frequency 
\begin{equation}
\label{eq:wSTnut}
\begin{aligned}
\wST^{\rm nut}
=&\sqrt{2}\pi\cos^{2}(\tfrac{1}{2}(\theff^{I}-\theff^{S}))
\\
\times&(\cos(\xi_{\rm ST})(J_{13}+J_{23})+\sin(\xi_{\rm ST})(J_{13}-J_{23})),
    \end{aligned}
\end{equation}
where $\xi_{\rm ST}$ represents the mixing angle between the $\ket{S_{0}\beta^{'}}$ and $\ket{T_{0}\beta^{'}}$ state
\begin{equation}
\label{eq:xiST}
\begin{aligned}
\xi_{\rm ST}=\frac{1}{2}\arctantwo(-\frac{J_{12}}{2},\frac{J_{13}-J_{23}}{4} \cos(\theff^{I}-\theff^{S})).
    \end{aligned}
\end{equation}

Application of a rotating magnetic field with \mbox{$\wrot=\wSTORM$} causes the states $\ket{S_{0}\beta^{'}}$ and $\ket{T_{-1}\alpha^{'}}$ to approximately follow the dynamics of a two-level system (TLS).

Consider now starting from the density operator in equation \ref{eq:rho0}. The spin-state populations at time $\tau$ under the TLS approximation are given by~\cite{messiah_albert_quantum_1962}

\begin{align}
\label{eq:oscil}
    &\bra{S_0\beta^{'}}\,\rho(\tau)\,\ket{S_0\beta^{'}} \simeq
        \hf\left(1 + \cos(\wSTnut \tau)
        \right),
\nonumber\\
    &\bra{T_{-1}\alpha^{'}}\,\rho(\tau)\,\ket{T_{-1}\alpha^{'}} \simeq
        \hf\left(1 - \cos(\wSTnut \tau)
        \right).
\end{align}
A complete population inversion between the $\ket{S_{0}\beta^{'}}$ and $\ket{T_{-1}\alpha^{'}}$ state may be achieved by applying the STORM pulse for a duration $\tau^{*}=\pi/\wSTnut$. Since the STORM pulse is only resonant with the $\ket{S_{0}\beta^{'}}\leftrightarrow\ket{T_{-1}\alpha^{'}}$ transition all other states experience negligible evolution. The idealised density operator after such a STORM pulse is thus given by
\begin{equation}
\begin{aligned}
\rho(\tau^{*})
\simeq&\frac{1}{2}\left(\ket{S_{0}\alpha^{'}}\bra{S_{0}\alpha^{'}}+\ket{T_{-1}\alpha^{'}}\bra{T_{-1}\alpha^{'}}\right)
\\
=&\mathds{1}/8+\cos(\theff^{S}) S_{z}/4+{\rm orth.}\;{\rm operators},
    \end{aligned}
\end{equation}
which indicates the generation of heteronuclear $S$ spin magnetisation proportional to $\cos(\theff^{S})$.
The state of the $S$ spins may be extracted by tracing over the $I$ spins of the system
\begin{equation}
\begin{aligned}
\rho_{S}(\tau^{*})={\rm Tr}_{I}\{\rho(\tau^{*})\}=&\mathds{1}/2+\cos(\theff^{S}) S_{z}.
    \end{aligned}
\end{equation}
As a result, the heteronuclear spins become fully polarised as $\cos(\theff^{S})$ approaches unity. Representing the transfer amplitude in a slightly more intuitive form:
\begin{equation}
\begin{aligned}
\cos(\theff^{S})=\frac{\wLS+\wrot}{\sqrt{(\wRS)^{2}+(\wLS+\wrot)^{2}}},
    \end{aligned}
\end{equation}
one may see that this condition is met if the sum of the bias $S$ spin Larmor frequency and angular rotation frequency exceed the $S$ spin nutation frequency ($\vert\wLS+\wrot\vert\gg\vert\wRS\vert$).

Although much of the discussion above has focused on the $\{\ket{S_{0}\beta^{'}},\ket{T_{0}\alpha^{'}}, \ket{T_{-1}\alpha^{'}}\}$ manifold, similar results hold for the $\{\ket{S_{0}\alpha^{'}},\ket{T_{0}\beta^{'}}, \ket{T_{+1}\beta^{'}}\}$ manifold, only difference being that the relevant energy difference has to be replaced by  
\begin{equation}
\begin{aligned}
\Delta E_{(-)}
&=\weff^{I}-\weff^{S}-2\pi\left(J_{12}-\frac{J_{13}+J_{23}}{4}\cos(\theff^{I}-\theff^{S})\right),
    \end{aligned}
\end{equation}
where the $(-)$ indicates that this transition will generally lead to negative heteronuclear magnetisation.

\section{\label{sec:Methods}Methods}
\subsection{\label{sec:Materials}Materials}
The precursor solution for fumarate was prepared by dissolving 100~mM disodium acetylene dicarboxylate, 100~mM sodium sulfite, and 6~mM \RuCat (CAS number: 99604-67-8) in D$_2$O, heating to 60\degC, and passing through a Millex\textregistered\ 0.22~$\mu$m PES filter.

\Para-hydrogen was  produced by passing hydrogen gas over an iron oxide catalyst packed in a 1/4 inch stainless steel tube cooled by liquid nitrogen which results in \para-enrichment level of 50\%. 

About 2\% of the fumarate molecules contain a naturally-occurring \Cth nucleus. The two \Hone nuclei and the \Cth nucleus form a three-spin-1/2 system of the type discussed above. The $J$-coupling parameters for the  molecular system are consistent with reference~\citenum{bengs_robust_2020,dagys_low-frequency_2021}.

\subsection{\label{sec:Equipment}Equipment}

\begin{figure*}[h!]
\centering
\includegraphics[width=0.8\textwidth]{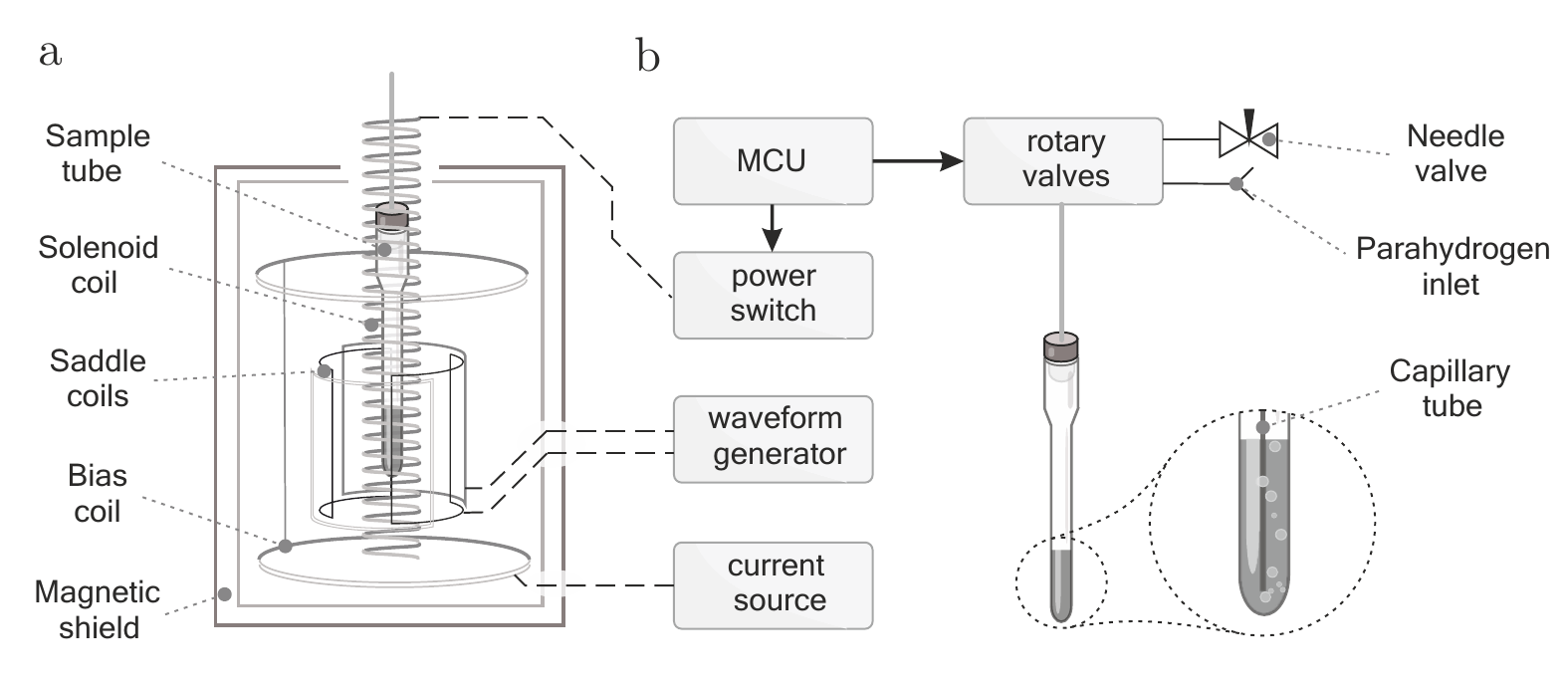}
\caption{\label{fig:setup}
Schematic diagram of the experimental setup. (a) Mu-metal shield and associated components. During the STORM pulse, the Helmholtz coil generates the bias field $\Bbias$ whereas two saddle coils produce the rotating field $\BSTORM$. (b) Gas-handling apparatus including a pressure NMR tube equipped with a capillary for bubbling of the \para-enriched \Htwo gas. MCU - micro-controller unit.
}
\end{figure*}
A sketch of the equipment is shown in figure~\ref{fig:setup}. 
The hydrogen gas is bubbled through the solution using a 1/16 inch PEEK capillary tube inserted inside a thin-walled Norell\textregistered\ pressure NMR tube. The Arduino Mega 2560 micro-controller board was used to actuate the Rheodyne MXP injection valves as well as a power switch connected to the solenoid coil. The 50~cm long and 15~mm wide coil was designed to provide a 170~$\mu$T field piercing through the TwinLeaf~MS-4 mu-metal shield. The rotating magnetic field was generated by two 30~cm long orthogonal saddle coils using a Keysight 33500B waveform generator with two channels synchronised with phase difference of $\pm90^\circ$.
The bias field was generated by the built-in Helmholtz coil of the Twinleaf shield, powered by a Keithley 6200 DC current source. 

\subsection{\label{sec:Procedure}Experimental Procedure}

\begin{figure}[h!]
\centering
\includegraphics[width=0.42\textwidth]{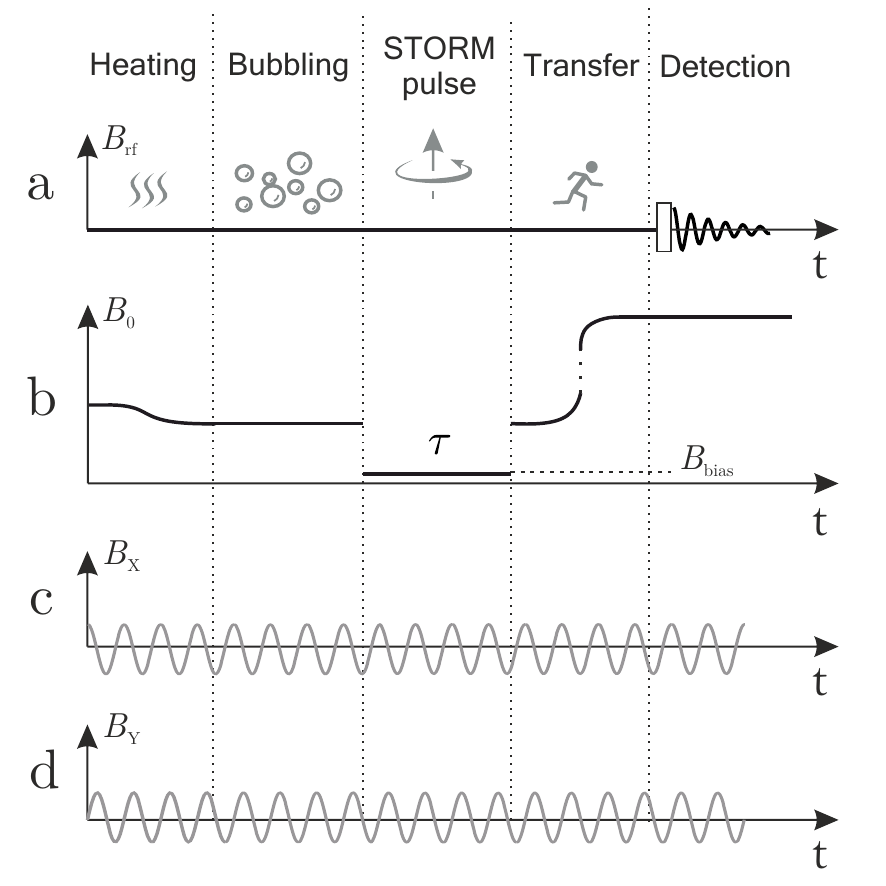}
\caption{\label{fig:sequence}
Detailed timing diagram for the STORM procedure. 
(a) A radiofrequency \Cth pulse is applied in high magnetic field at the end of the procedure to initiate signal acquisition. 
(b) Magnetic field profile along the $z$-axis, showing the ambient laboratory field, the change in field as the sample is placed in the shield, the bias field $\Bbias$ during the STORM pulse, the removal of the sample from the shield and insertion into the high-field NMR magnet.
(c) Magnetic field generated along the X direction and (d) Y direction oscillating with frequency $\wSTORM$ and peak amplitude $\BSTORM$. The oscillating fields are generated constantly, but only resonant for the time period $\tau$.
}
\end{figure}

Figure~\ref{fig:sequence} gives an overview of the experimental protocol including the magnetic field experienced by the sample as a function of time. Each experiment starts by heating 250\uL of the sample mixture to $\sim90 \degC$ in the ambient magnetic field of the laboratory ($\sim$ 110~$\mu$T), followed by insertion into the magnetic shield where a solenoid coil generates a magnetic field of similar magnitude. \Para-enriched hydrogen gas is bubbled through the solution at 6 bar pressure for 30~seconds. The rotating field is generated constantly by the waveform generator, the amplitude is kept constant ($\BSTORM=4\uT$). The solenoid is switched off by a relay for a period $\tau$ to pre-set the bias field ($\Bbias$) and to mimic a STORM pulse. Afterwards the sample was removed manually and inserted into the Oxford 400~MHz magnet equipped with a Bruker Avance Neo spectrometer.

The \Cth free-induction decays were initiated by a hard pulse of 14.7~kHz rf amplitude and recorded with 65~k point density at a spectral width of ~200~ppm. Additional \Hone decoupling was used for all experiments. Thermal equilibrium \Cth spectra were recorded at room temperature with a recycle delay of 120~s, averaging the signal over 512 transients.

\section{\label{sec:results}Results}

\begin{figure*}[h!]
\centering
\includegraphics[width=0.8\textwidth]{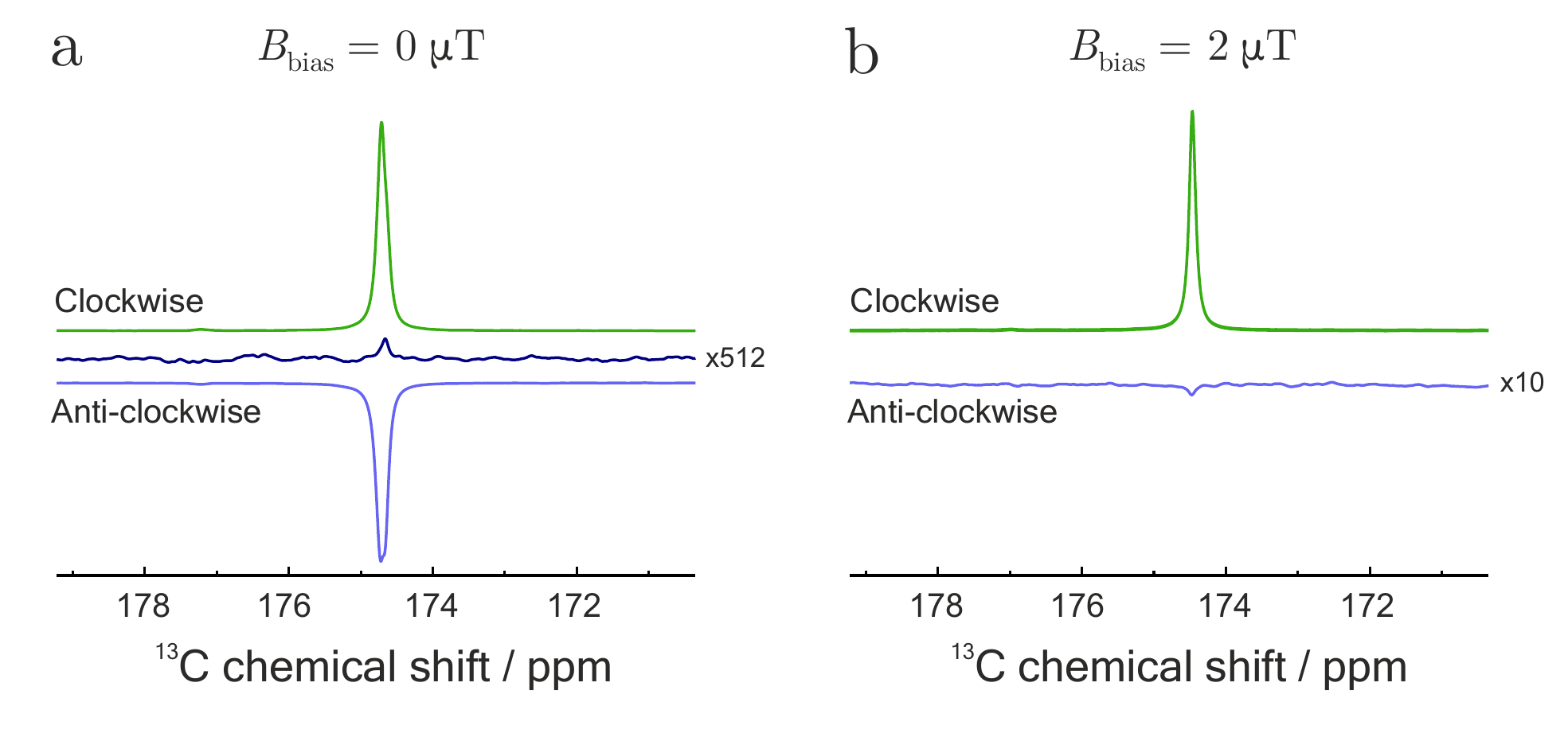}
\caption{\label{fig:spectra}  
\Hone-decoupled \Cth spectra of fumarate at a field of 9.41~T. Samples subjected to clockwise (green lines) and anti-clockwise (blue lines) rotating magnetic fields with a peak amplitude of 4~$\mu$T. (a) STORM polarisation transfer at zero-field ($\Bbias\simeq 0$ T) and rotation frequency of $\wSTORM=1150$~Hz. (b) STORM polarisation transfer inside a small bias field ($\Bbias\simeq 2\;\mu{\rm T}$) and rotation frequency of $\wSTORM=223$~Hz. The black trace in (a) corresponds to a \Cth NMR spectrum at thermal equilibrium averaged over 512 transients.}
\end{figure*}

Figure~\ref{fig:spectra} shows single-transient hyperpolarized \Cth NMR spectra, obtained using the STORM procedure as a function of the rotational direction in the presence of a 0~T bias field (figure~\ref{fig:spectra}(a)) and a 2~$\mu$T bias field (figure~\ref{fig:spectra}(b)). In both cases the STORM pulse duration was set to 0.2~s with a peak amplitude of $\BSTORM=4~\mu$T. The rotation frequencies, 1150 Hz for (a) and 223 Hz for (b), correspond to the root of the equation~\ref{eq:delta_E}.
The observed \Cth polarization levels clearly depend upon the sense of rotation. At zero field the relevant LAC conditions for positive and negative \Cth magnetization are centred symmetrically around a zero rotation frequency. A simple inversion of the rotation frequency therefore enables selection of either positive and negative \Cth magnetization.
In the presence of a non-vanishing bias field the symmetry with respect to the rotation frequency is broken. This means that a single frequency can only match one condition and no signal is observed when the rotation is reversed (see Fig.~\ref{fig:spectra}(b)).

\begin{figure}[]
\centering
\includegraphics[width=0.45\textwidth]{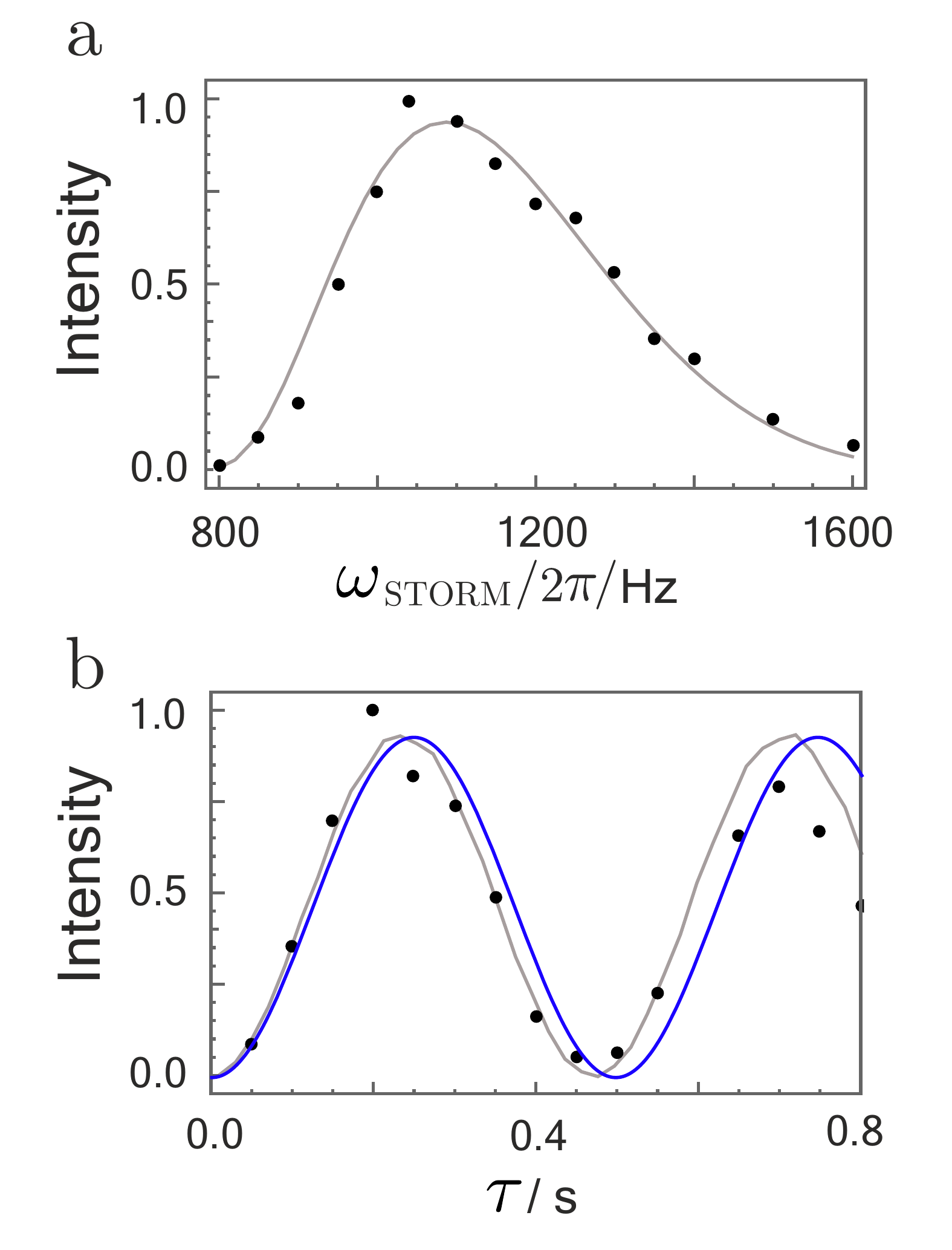}
\caption{\label{fig:zf_profiles} 
(a) Hyperpolarized fumarate intensities as a function of STORM pulse frequency $\wSTORM$ \blue{with a fixed pulse duration of 200~ms}. (b) Hyperpolarized fumarate intensities as a function of pulse duration $\tau$ \blue{with a fixed STORM frequency of 1100~Hz.} The bias field and rotating field are fixed at 0~$\mu$T and 4~$\mu$T, respectively. Grey lines represent numerical \emph{SpinDynamica}~\cite{bengs_spindynamica_2018} simulations, whereas blue lines plot the analytical solution given by equation~\ref{eq:oscil}. The intensity scales are normalised to the maximum signal obtained.}
\end{figure}
\begin{figure}[]
\centering
\includegraphics[width=0.45\textwidth]{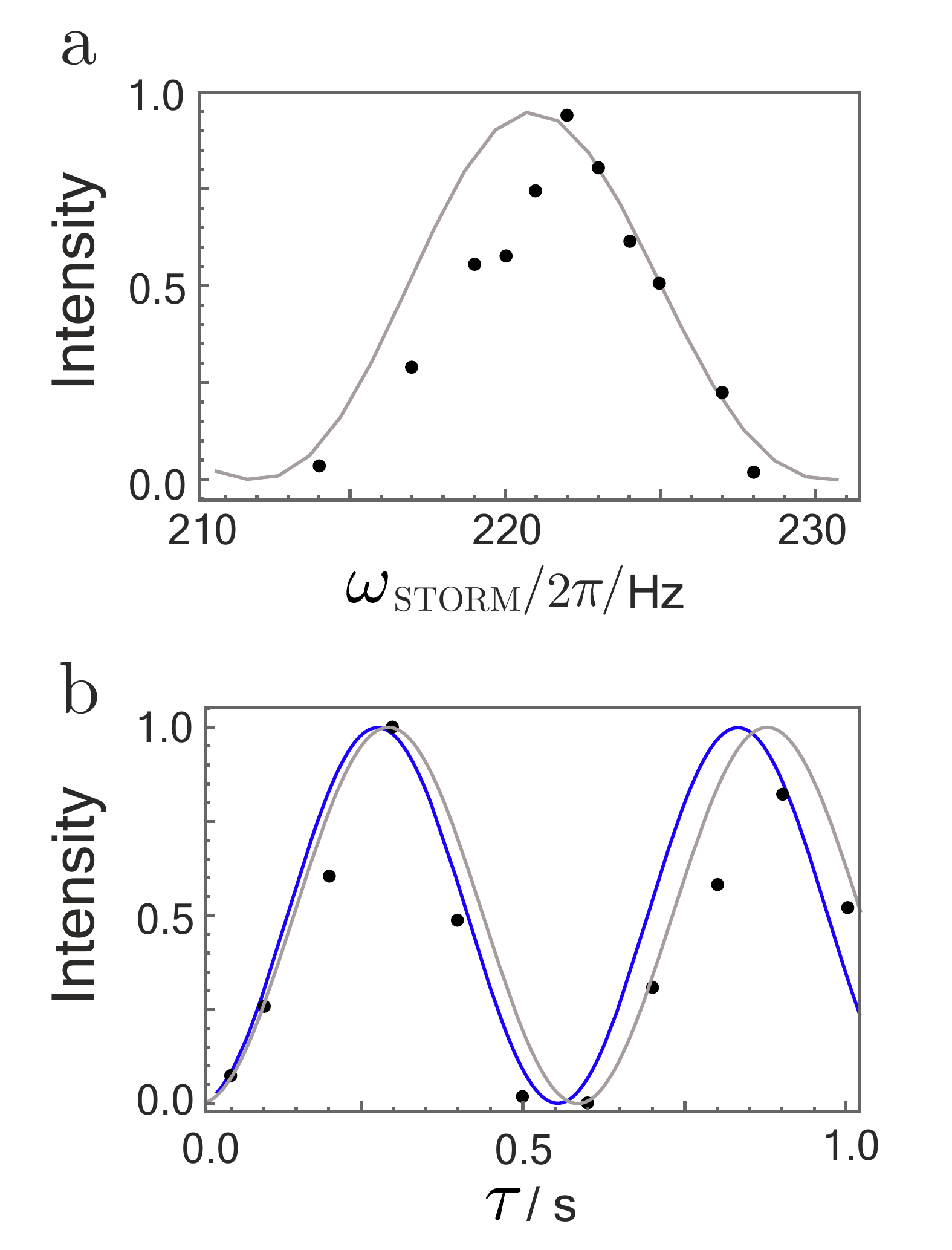}
\caption{\label{fig:profiles} 
(a) Hyperpolarized fumarate intensities as a function of STORM pulse frequency $\wSTORM$ \blue{with a fixed pulse duration of 400~ms}. (b) Hyperpolarized fumarate intensities as a function of pulse duration $\tau$ \blue{with a fixed STORM frequency of 222~Hz.} The bias field and rotating field are fixed at 2~$\mu$T and 4~$\mu$T, respectively. Grey lines represent numerical \emph{SpinDynamica}~\cite{bengs_spindynamica_2018} simulations, whereas blue lines plot the analytical solution given by equation~\ref{eq:oscil}. The intensity scales are normalised to the maximum signal obtained.
}
\end{figure}

The solid black line in figure~\ref{fig:spectra}(a) represents a reference \Cth spectrum averaged over 512 transients. The spectrum was obtained on the hydrogenated sample after thermal equilibration. Comparison of these spectra allows an estimation of the \Cth polarization levels, which in this case corresponds to $p^{S}_{z}\simeq6\%$. These results are comparable with previous methods under similar experimental conditions.~\cite{eills_polarization_2019,rodin_constant-adiabaticity_2021,rodin_constant-adiabaticity_2021-1,dagys_low-frequency_2021} Significant improvements in the polarisation are expected by addressing few aspects of the setup. Fully enriched \para-hydrogen would lead to 3-fold enhancement, whereas careful optimization of the reaction conditions would further lead to a better polarization yield\cite{eills_polarization_2019,rodin_constant-adiabaticity_2021}. Some minimal losses could be also avoided by using a fully automated experimental procedure. 

Integrated \Cth signal amplitudes as a function of the rotation frequency and the STORM pulse duration $\tau$ at bias field of 0~T and 2~$\mu$T are shown in figures~\ref{fig:zf_profiles}~and~\ref{fig:profiles}. Each experimental point was obtained from a separate experiment on a fresh sample. The experimental data has been normalised to unity to enable a qualitative comparison with numerical simulations and analytically derived curves based on equation~\ref{eq:oscil}. The agreement between both curves and the experimental data is gratifying.

The frequency profiles obtained at different bias fields display a significant change in their width. 
At zero-field the full width at half maximum of the profile was estimated to be $\sim$350~Hz whereas at 2~$\mu$T bias field width got reduced to $\sim$5~Hz.
These matching conditions are much broader than the 0.4~Hz width observed in the profiles using the WOLF method~\cite{dagys_low-frequency_2021}. In agreement with the analytic expression given by equation~\ref{eq:wSTnut}, the polarisation transfer rate did not vary dramatically with an increase in the bias field. For both cases, equation~\ref{eq:wSTnut} correctly predicts the polarisation transfer rate to be approximately $\sim 2$~Hz, where we have used the \textit{J}-coupling parameters of fumarate given in refs~\citenum{bengs_robust_2020,dagys_low-frequency_2021}.

\section{Conclusions}

In this work we have studied the polarization transfer from singlet order to heteronuclear magnetization in the context of \para-hydrogen induced polarisation. The polarisation transfer is achieved through a combination of a rotating magnetic field and a small bias field. 
Despite low values of the static bias field, the rotation frequency needed to drive the transfer is strikingly large. However, the principle of the method described here is rather intuitive and a simple explanation may be given within the framework of level-anti crossings, at least when the Hamiltonian is expressed within the interaction frame of the rotating magnetic field. Based on the LAC picture we were able to establish the resonance conditions for singlet-triplet mixing and determine the corresponding population transfer rate. In particular, we have shown that the rotational direction plays an important role in correctly establishing the resonance condition, and should be chosen carefully in the experimental context.

There are other methods to convert nuclear singlet order into heteronuclear polarization, including resonant pulse schemes in high field as well as magnetic field-cycling at low magnetic fields~\cite{johannesson_transfer_2004,theis_microtesla_2015,theis_light-sabre_2014,devience_homonuclear_nodate,rodin_constant-adiabaticity_2021-1,rodin_constant-adiabaticity_2021,cavallari_effects_2015,eills_polarization_2019,knecht_rapid_2021,dagys_low-frequency_2021,dagys_nuclear_2021,bengs_robust_2020}. The STORM method introduced here is conceptually simple and \blue{provides a  few advantages} over other existing low field methods. The potential polarization losses caused by additional relaxation effects in high magnetic fields are entirely avoided with use of low magnetic fields~\cite{rodin_constant-adiabaticity_2021-1,knecht_mechanism_2018, knecht_indirect_2019}. \blue{However, at ultra-low fields quadrupolar nuclei such as $^2$H or $^{14}$N often act as polarization sinks, and may lead to a significant drops in the polarization transfer efficiency.}
\blue{The presence of quadrupolar nuclei is expected to be particularly disrupting to adiabatic field sweep methods, some of which are routinely utilised in the generation of hyperpolarised \mbox{(1-\Cth)fumarate}~\cite{rodin_constant-adiabaticity_2021,rodin_constant-adiabaticity_2021-1,knecht_rapid_2021}.}
\blue{In contrast to field sweep methods, the STORM method allows one to freely select the strength of magnetic fields as well as the rotation frequency.} \blue{It is thus conceivable that optimal conditions for the magnetic field strength and rotation frequency exist at which the $^2$H or $^{14}$N spins do not interfere with the polarisation transfer process.}
We therefore believe that STORM pulses represent promising candidates for a new class of quadrupolar decoupled polarization transfer methods in the near future. Applications to other hyperpolarization techniques such as PHIP-SABRE (Signal Amplification by Reversible Exchange) are also conceivable~\cite{adams_reversible_2009,theis_light-sabre_2014,theis_microtesla_2015,knecht_mechanism_2018}.

\section*{Conflicts of interest}
There are no conflicts to declare.

\section*{Acknowledgements}
We acknowledge funding received by the Marie Skłodowska-Curie program of the European Union (grant number 766402), the European Research Council (grant 786707-FunMagResBeacons), and EPSRC-UK (grants EP/P009980/1, EP/P030491/1, EP/V055593/1). We thank Malcolm H. Levitt for valuable input and support during preparation of the manuscript.

\section*{Appendix}
\label{app:STORMnut}
\subsection*{STORM nutation frequency}

Consider the manifold defined by
\begin{equation}
\begin{aligned}
V_{1}=\{\ket{S_{0}\beta^{'}},\ket{T_{0}\alpha^{'}}, \ket{T_{-1}\alpha^{'}}\},
\end{aligned}
\end{equation}
the argument is analogous for the other manifold. The matrix representation of $\tilde{H}$ restricted to $V_{1}$ manifold is of the following form
\begin{strip}
\begin{equation}
\begin{aligned}
{[}\tilde{H}{]}_{V_{1}}=
\left[\begin{array}{ccc}
     -\frac{1}{2}(\weff^{S}+3\pi J_{12})
     & \frac{\pi}{2}\cos(\theff^{I}-\theff^{S})(J_{13}-J_{23})
     &
     \blacksquare
     \\
     \frac{\pi}{2}\cos(\theff^{I}-\theff^{S})(J_{13}-J_{23}) & -\frac{1}{2}(\weff^{S}-\pi J_{12})
     & 
     \blacksquare
     \\
     \blacksquare
     &
     \blacksquare
     &
     \blacksquare
\end{array}\right],
    \end{aligned}
\end{equation}
\end{strip}
where the black squares indicate non-zero, but irrelevant matrix elements. 

The energy separation between the $\ket{S_{0}\beta^{'}}$ and $\ket{T_{0}\beta^{'}}$ state equals $\vert2\pi J_{12}\vert$ and is not quite sufficient to perform a TLS approximation. We thus diagonalise the corresponding subspace making use of the mixing angle

\begin{equation}
\begin{aligned}
\xi_{\rm ST}=\frac{1}{2}\arctantwo(-\frac{J_{12}}{2},\frac{J_{13}-J_{23}}{4} \cos(\theff^{I}-\theff^{S})).
    \end{aligned}
\end{equation}

After the diagonalisation process the matrix representation of $\tilde{H}$ takes the form
\begin{strip}
\begin{equation}
\begin{aligned}
{[}\tilde{H}{]}_{V_{1}}=
\left[\begin{array}{ccc}
    \blacksquare
     & 0
     &
     \frac{1}{2}\wSTnut
     \\
     0
     & 
     \blacksquare
     & 
     \blacksquare
     \\
     \frac{1}{2}\wSTnut
     &
     \blacksquare
     &
     \blacksquare
\end{array}\right]\xrightarrow{\rm TLS\; approximation}\left[\begin{array}{cc}
    \blacksquare
     &
     \frac{1}{2}\wSTnut
     \\
     \frac{1}{2}\wSTnut
     &
     \blacksquare
\end{array}\right],
    \end{aligned}
\end{equation}
\end{strip}
where $\wSTnut$ is given by equation \ref{eq:wSTnut}.
\clearpage





\bibliography{RotWOLF.bib} 
\bibliographystyle{rsc} 

\end{document}